# Nuclear Quadrupole Hyperfine Structure in HC$^{14}$N/H$^{14}$NC and DC$^{15}$N/D$^{15}$NC Isomerization: A Diagnostic Tool for Characterizing Vibrational Localization

Bryan M. Wong*

Large-amplitude molecular motions which occur during isomerization can cause significant changes in electronic structure. These variations in electronic properties can be used to identify vibrationally-excited eigenstates which are localized along the potential energy surface. This work demonstrates that nuclear quadrupole hyperfine interactions can be used as a diagnostic marker of progress along the isomerization path in both the HC$^{14}$N/H$^{14}$NC and DC$^{15}$N/D$^{15}$NC chemical systems. *Ab initio* calculations at the CCSD(T)/cc-pCVQZ level indicate that the hyperfine interaction is extremely sensitive to the chemical bonding of the quadrupolar $^{14}$N nucleus and can therefore be used to determine in which potential well the vibrational wavefunction is localized. A natural bonding orbital analysis along the isomerization path further demonstrates that hyperfine interactions arise from the asphericity of the electron density at the quadrupolar nucleus. Using the CCSD(T) potential surface, the quadrupole coupling constants of highly-excited vibrational states are computed from a one-dimensional internal coordinate path Hamiltonian. The excellent agreement between *ab initio* calculations and recent measurements demonstrates that nuclear quadrupole hyperfine structure can be used as a diagnostic tool for characterizing localized HCN and HNC vibrational states.

## I. Introduction

The H–C≡N: ⇌ H–N≡C: isomerization on the ground electronic potential energy surface exemplifies one of the simplest bond-breaking processes in molecules. These high-barrier isomerization processes are of fundamental importance to many areas of chemistry since they involve the breaking and formation of bonds, a highly energetic process found in combustion and interstellar plasmas.[1] In particular, investigations of the HCN ⇌ HNC isomerization system have an additional importance because they allow the computation of [HCN]/[HNC] abundance ratios, which provide further insights on the composition and chemistry of dark interstellar clouds.[2,3] For these reasons, numerous spectroscopic studies have been carried out on HCN[4-9], with the intent to directly observe its isomerization to HNC and to confirm theoretical calculations.[10-17]

A primary motivation for studying the HCN ⇌ HNC isomerization is to observe vibrationally-excited bending states which lie close to the energy barrier maximum. These highly-excited vibrational states are of particular interest because they have wavefunction amplitude localized along the minimum energy path and therefore reveal electronic features of the potential that control intramolecular dynamics. In the case of chemical isomerization, bonds are broken and new bonds are formed, and the electronic wavefunction becomes severely deformed from that of the equilibrium nuclear configuration. In light of this observation, the change in electronic structure along the HCN ⇌ HNC isomerization path can be used as an effective diagnostic for identifying delocalized vibrational states between the HCN and HNC regions of the potential. Indeed, Bowman et al. have already performed three-dimensional *ab initio* calculations on the dipole moment and showed that isomerizing vibrational states possess significantly smaller dipole moments than localized HCN or HNC states.[18] Fig. 1 depicts the HCN ⇌ HNC isomerization path as a function of the ∠HCN Jacobi angle. At the endpoints of this path, the HCN and HNC *a*-axis dipoles are nearly equal in magnitude but opposite in sign, with the *a*-axis dipole moment function changing sign near the transition state. Since delocalized vibrational eigenfunctions near the transition state sample both HCN and HNC configurations, their dipole-moment expectation values are nearly zero. Consequently. Stark effect measurements of dipole moments in vibrationally-excited states are direct experimental observables related to the extent of delocalization.

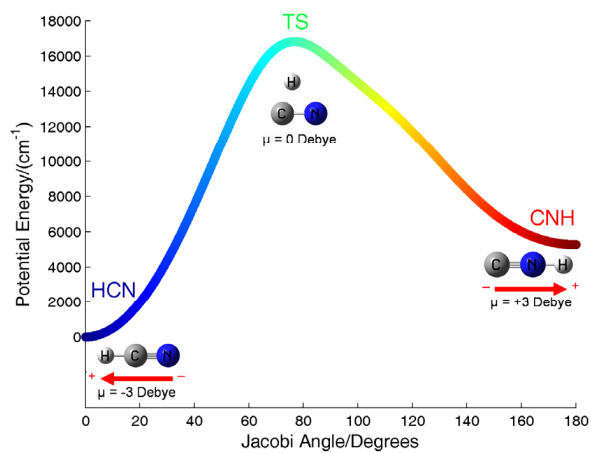

**Fig. 1** One-dimensional potential energy surface for the HCN ⇌ HNC isomerization as a function of the Jacobi angle. The *a*-axis dipole moment changes sign from -3 Debye to +3 Debye along the HCN (0°) → HNC (180°) isomerization path. The relaxed potential energy curve was obtained from the CCSD(T)/cc-pCVQZ level of theory.

Another electronic property useful in identifying isomerization states is nuclear quadrupole hyperfine interactions. Hyperfine structure originates from the nuclear quadrupole moment of a spin $I \geq 1$ nucleus interacting with the electric field gradient due to the electronic wavefunction. For both $HC^{14}N$ and $H^{14}NC$, only the $^{14}N$ nitrogen nucleus possesses a nuclear quadrupole moment, and the nuclear quadrupole coupling constants are highly sensitive to which isomer is measured. Qualitatively, the change in nuclear quadrupole hyperfine structure for $HC^{14}N/H^{14}NC$ is due to variations in the distances and angles between atoms bonded with the $^{14}N$ nucleus. Consequently, hyperfine structure measurements provide another probe to identify the onset of delocalization.

In this work, the concept of using nuclear quadrupole couplings for identifying vibrational states is explained in detail to complement and extend the experimental data reported in a recent communication.[19] A description of nuclear quadrupole structure is presented, and a detailed account of the computational steps involved in obtaining the current results is provided. Other diagnostic methods such as Weinhold's natural bond orbital analysis[20] are also computed to determine the various interactions which cause changes in hyperfine structure. In addition, new hyperfine calculations applied to the isotopically-substituted species $DC^{15}N$ and $D^{15}NC$ are also analyzed. These calculations allow a useful comparison to experimental data and reveal mechanistic details for the evolution of vibrational character.

## II. Quadrupole Coupling Constants of Nuclei

In a molecule, the electric field gradient $\mathbf{q_J}$ at the site of a nucleus $J$ is given by the second derivatives of the potential $V$ with respect to the Cartesian coordinates. Accordingly, $\mathbf{q_J}$ is a symmetric second rank tensor with Cartesian components $q_{xx,J} = \partial^2 V/\partial x^2$, $q_{xy,J} = \partial^2 V/\partial x \partial y$, etc. Given the electronic wavefunction, $\psi$, the electric field gradient component at a particular nucleus $J$, $q_{xy,J}$, is given by a sum of nuclear and electronic terms[21]:

$$q_{xy,J} = q_{xy,J}^{\text{nucl}} + q_{xy,J}^{\text{elec}}$$
$$= e\sum_{I \neq J} \frac{Z_I \left(3x_{IJ}y_{IJ} - r_{IJ}^2\right)}{r_{IJ}^5} - e\langle\psi|\sum_i \frac{3x_{iJ}y_{iJ} - r_{iJ}^2}{r_{iJ}^5}|\psi\rangle, \quad (1)$$

where $e$ is the charge of an electron, $Z_I$ is the charge on nucleus $I$, and $i$ labels the electrons. The coordinate system in which the Cartesian form of the tensor is diagonal is the principal axis system. From eqn (1), the Cartesian tensor is traceless ($q_{xx} + q_{yy} + q_{zz} = 0$), and therefore, only two components are independent in the principal axis system. These two independent components are conventionally taken to be $q_{zz}$ and the asymmetry parameter, $\eta = \left(q_{xx} - q_{yy}\right)/q_{zz}$.[21] The principal component, $q_{zz}$, follows from simplification of eqn (1):

$$q_{zz,J} = e\sum_{I \neq J} \frac{Z_I \left(3z_{IJ}^2 - r_{IJ}^2\right)}{r_{IJ}^5} - e\langle\psi|\sum_i \frac{3z_{iJ}^2 - r_{iJ}^2}{r_{iJ}^5}|\psi\rangle. \quad (2)$$

The quantity determined from experimental data is the nuclear quadrupole coupling constant $\chi_{zz}$, which is usually measured in units of MHz and directly proportional to $q_{zz}$:

$$\chi_{zz} = eQq_{zz}/h, \quad (3)$$

where $e$ is the charge of an electron, $h$ is Planck's constant, and $Q$ is the scalar quadrupole moment of the nucleus. The nuclear quadrupole coupling is also characterized by the asymmetry parameter $\eta$, but since these values were not reported in the experimental study of Bechtel et al.[19], only the principal nuclear quadrupole coupling constant is presented in the remainder of this work. Qualitatitvely, the nuclear quadrupole moment describes the spheroidal shape of positive nuclear charge, and only nuclei with spin $I \geq 1$ have nonzero nuclear electric quadrupole moments.[22] The magnitude and sign of the nuclear quadrupole moment indicates the shape of the atomic nucleus. For example, a zero value of $eQ$ indicates a spherically symmetric charge distribution and no quadrupole moment. A positive sign indicates that the asymmetric distribution of protons is such that there is an elongation along the body $z$ axis, and a prolate spheroid results. Finally if the asymmetric distribution of protons is such that a flattening of the nucleus along the $x$ and $y$ axes occurs, an oblate spheroid is produced. Since the nuclear quadrupole coupling constant, $\chi$, is directly related to the asphericity of the electron density at the probe nucleus, the prediction of $\chi$ allows an estimation of the $p\pi$ and $p\sigma$ bonding within the molecule.

## III. Ab Initio Calculations

From the second term in eqn (2), the calculation of nuclear quadrupole coupling constants involves an expectation value of the electric field gradient at the nucleus. Due to the dependence on the electronic environment close to the nucleus of interest, an accurate calculation of quadrupole coupling constants puts strict demands on the basis set and also the level of electron correlation. In particular, other researchers have already shown that *ab initio* calculations using the correlation consistent basis sets of Dunning and coworkers[23], are not well-suited for accurate calculations of quadrupole coupling constants.[24] As it is the region close to the nucleus of interest that needs to be accurately described, the use of large core-valence basis sets is necessary.[2]

For the HCN $\rightleftarrows$ HNC isomerization, all calculations were performed at the coupled cluster level of theory by including single and double excitations together with a perturbative treatment of triple excitations (CCSD(T)). The relaxed geometry parameters were obtained with the core-valence cc-pCVTZ basis set, and all core electrons were correlated at the CCSD(T) level to account for core correlation effects. CCSD(T) single-point energies and electric field gradients were subsequently performed with a larger cc-pCVQZ basis set at the CCSD(T)/cc-pCVTZ optimized geometries. The HCN ⇌ HNC quadrupole coupling constants on the one-dimensional isomerization path were calculated by choosing a grid of 24 values of the Jacobi angle between the HCN ($\theta = 0°$) and HNC ($\theta = 180°$) isomers and optimizing all other internal coordinates to minimize the total energy. The Jacobi angle used in the present work is defined as the angle between the two Jacobi vectors $r$ and $R$, where $r$ is the N–C displacement vector, and $R$ is the displacement vector between the C–N center of mass and the H atom. All CCSD(T) calculations were performed using the ACES II (Mainz-Austin-Budapest Version) set of programs[25] with analytic gradients for both geometry and electric field gradient calculations.

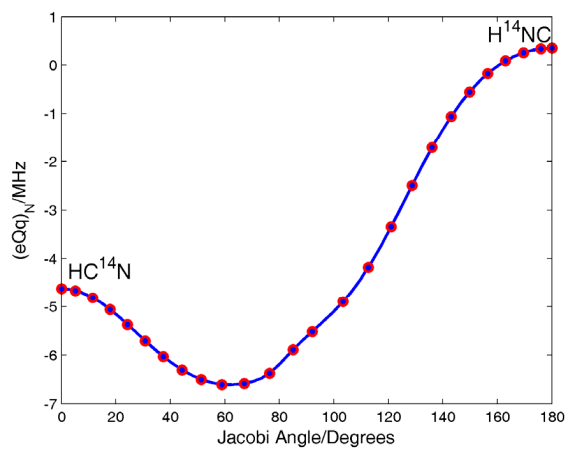

**Fig. 2** Dependence of the nuclear quadrupole coupling constant $(eQq)_N$ along the relaxed $HC^{14}N \rightleftarrows H^{14}NC$ isomerization path. The quadrupole coupling constant does not follow a simple linear trend along the optimized path.

## IV. Results and Discussion

### A. Quadrupole Coupling Constants

The principal component of the quadrupole coupling constant tensor, $(eQq)_N$, along the optimized isomerization path from $HC^{14}N$ to $H^{14}NC$ is shown in Fig. 2. For $HC^{14}N$, the quadrupole coupling constant is large and negative, $(eQq)_N = -4.62$ MHz, whereas for $H^{14}NC$, it is small and positive $(eQq)_N = 0.353$ MHz. It should be noted that the most recent *ab initio* calculations by Pd and Chandra[26] on $H^{14}NC$ also predict a small nuclear quadrupole coupling constant $-313$ kHz $\geq (eQq)_N \geq -288$ kHz, but of the wrong sign. Interestingly, Fig. 2 shows that the $(eQq)_N$ values do not follow a simple linear trend along the isomerization path. Instead, the magnitude of the coupling constant, $|(eQq)_N|$, initially increases as a function of the Jacobi angle up to $\theta = 76°$, which corresponds to the transition state geometry. Once the Jacobi angle is increased past the transition state structure, $|(eQq)_N|$ decreases until it is approximately zero.

In order to explain these trends, a natural bond orbital (NBO) analysis[20] was utilized to calculate electron populations of the C and N atoms during the isomerization process. Using the density matrix of the CCSD wavefunction, the NBO procedure yields a set of localized orbitals which give the most accurate Lewis-like description of the total electron density. The NBO electron populations of the N and C atoms as a function of the Jacobi angle are plotted in Figs. 3 (a) and (b). According to the model of Townes and Dailey[27], the primary contribution to the electric field gradient, $q$, is due to an unequal distribution of electrons among the $p_x$, $p_y$, and $p_z$ orbitals of the quadrupolar nucleus. In the HCN isomer ($\theta = 0°$), the N nucleus has 1.47 electrons in a $p_z$ orbital which are used to form both a C–N $\sigma$ bond and a portion of the N: lone pair. The $p_x$ and $p_y$ orbitals on the N nucleus are equally filled with 1.08 electrons which contribute to the two $\pi$ bonds with the C nucleus. The excess of electrons in the $p_z$ orbital creates a molecular electric field which is highly anisotropic and results in a large quadrupole coupling constant. As the Jacobi angle is increased, the NBO analysis indicates the initially-symmetric electron density in the $xy$-plane is broken as the $p_x$ and $p_y$ orbitals also become unequally filled. This further unequal distribution of electrons creates an even larger asphericity of electron density and causes $|(eQq)_N|$ to increase. However, as the transition state geometry is approached ($\theta = 76°$), the *difference* between the $p_x$ and $p_y$ electron populations is maximal, and $|(eQq)_N|$ is largest in this region. In the post-transition-state region ($\theta > 76°$), the $p_z$ electron population decreases as the lone pair becomes shared to form the N–H bond. Finally once HNC is formed ($\theta = 180°$), the NBO analysis shows that the $p_x$, $p_y$, and $p_z$ orbitals are all equally filled (~ 1.43 electrons), and the quadrupole coupling constant is approximately zero.

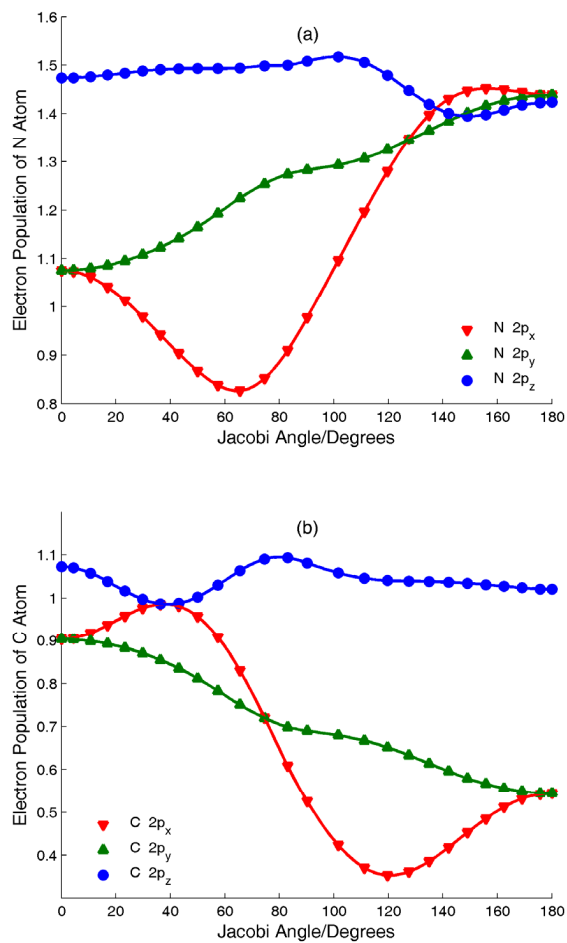

**Fig. 3** Natural Bond Orbital *p*-electron populations computed for (a) the nitrogen atom and (b) the carbon atom as a function of the ∠HCN Jacobi angle. In Fig. 3 (a), the nitrogen electron density near $\theta = 76°$ is highly aspherical since the distribution of electrons among the $p_x$, $p_y$, and $p_z$ orbitals is maximally unequal. Conversely, the $p_x$, $p_y$, and $p_z$ orbitals are nearly equally filled once the HNC isomer at $\theta = 180°$ is formed. In Fig. 3 (b), the *p*-electron population of the carbon atom approximately follows the mirror image of the nitrogen electron population depicted in Fig. 3 (a).

A nearly identical chemical system which exhibits completely different nuclear quadrupole hyperfine structure is the $D-C\equiv{}^{15}N: \rightleftarrows D-{}^{15}N\equiv C:$ isomerization. In these isotopomers, only the deuteron has a quadrupole moment ($I_D = 1$) while both the ${}^{15}N$ and ${}^{12}C$ nuclei do not $(I_{{}^{15}N} = I_{{}^{12}C} = 1/2)$. As a result, the $DC^{15}N \rightleftarrows D^{15}NC$ isomerization provides a complementary description of quadrupole hyperfine structure from the alternate viewpoint of only the deuteron nucleus. Fig. 4 shows the quadrupole coupling constant $(eQq)_D$ along the relaxed isomerization path from $DC^{15}N$ to $D^{15}NC$. In contrast to the $HC^{14}N \rightleftarrows H^{14}NC$ isomerization, the quadrupole coupling constants for both $DC^{15}N$ and $D^{15}NC$ are small, positive, and nearly equal: $(eQq)_{DC^{15}N} = 0.212$ MHz and $(eQq)_{D^{15}NC} = 0.279$ MHz. The striking differences between Figs. 2 and 4 can also be explained in terms of the chemical environment experienced by the quadrupolar nucleus. Since the quadrupole coupling, $(eQq)_D$, arises from the interaction with the nuclei to which the deuteron is bound, it reflects the anisotropy of the molecular electric field at the position of the probe nucleus. Therefore, if the nucleus is located in a highly symmetric chemical environment, the electric field gradient is small. Near the transition state, the deuteron is positioned close to the center of the $C–{}^{15}N$ bond and experiences a fairly symmetrical field (the electronegativities of both C and N are fairly equal), and the electric field gradient is nearly zero as shown in Fig. 4. Consequently, a plot of $(eQq)_D$ as a function of Jacobi angle is nearly symmetrical about the transition state point.

**B. Hamiltonian**

In order to make quantitative comparisons with experimental data, the quadrupole coupling constants in Figs. 2 and 4 must be quantum-mechanically averaged over vibrational wavefunctions obtained from the one-dimensional isomerization potential. In the present work, an internal-coordinate Hamiltonian is used to determine the eigenvalues and eigenfunctions corresponding to the large-amplitude hydrogen motion. A detailed description of the internal coordinate path Hamiltonian and its derivation has been given by Tew et al.[28] Their parametrization of a large-amplitude motion with a single internal coordinate is based on the work of Hougen, Bunker, and Johns who were among the first workers to include large-amplitude motion in their semi-rigid bender model.[29] Other workers, such as Szalay and Nesbitt, have extended this approach to account for nonrigid effects of large-amplitude internal motion in other molecules.[30,31] The formulation of Tew et al. is closely related to the reaction path Hamiltonian by Miller, Handy, and Adams[32] with the exception that the internal coordinate path need not be exactly parallel to the minimum energy path. As reported previously[33], the following approximations allow for the computation to be manageable: (1) the inertia tensor depends weakly on the small-amplitude coordinates $Q_k$ ($k = 1, 2, …, 3N$-7), and only the terms in the inertia tensor that depend on the large-amplitude coordinate $s$ are retained; (2) the Coriolis terms are linear in the small-amplitude coordinates $Q_k$, and their contribution to the kinetic energy is neglected; (3) numerically enforcing the Eckart conditions minimizes many of the couplings between the large-amplitude motion and the overall rotation of the molecule. It follows from these approximations that the kinetic energy operator for the large-amplitude motion with total angular momentum $J = 0$ can be written in the following form (cf. eqn (10) of ref. 33)

$$\hat{T} = \frac{1}{2}\hat{p}_s I_{0ss}^{-1}\hat{p}_s + \frac{1}{2}\mu^{1/4}\left(\hat{p}_s I_{0ss}^{-1}\mu^{-1/2}\left(\hat{p}_s \mu^{1/4}\right)\right), \qquad (4)$$

where $\hat{p}_s (=-i\hbar\partial/\partial s)$ is the momentum conjugate to the large-amplitude coordinate $s$. Since the isomerization process corresponds to a bend-like motion, the most appropriate choice for the large-amplitude parameter $s$ is the Jacobi angle $\theta$. The scalar terms $I_{0ss}^{-1}$ and $\mu$ are given by

$$I_{0ss}^{-1}(s) = \left(\sum_{i=1}^{N}\mathbf{a}_i'(s)\cdot\mathbf{a}_i'(s)\right)^{-1}, \qquad (5)$$

$$\mu(s) = I_{0ss}^{-1}\cdot\det(\mathbf{I}_0^{-1}), \qquad (6)$$

and $\mathbf{I}_0$ is the normal $3\times 3$ Cartesian inertia tensor along the isomerization path. The vectors $\mathbf{a}_i \left(=m_i^{1/2}\mathbf{r}_i\right)$ are the mass-weighted Cartesian coordinates of the $i$th atom at a point on the path $s$ with respect to the Eckart axis system, and $\mathbf{a}_i' \left(=d\mathbf{a}_i/ds\right)$. Finally, the operator $\hat{p}_s$ operates only within the parentheses in eqn 4; that is, the next to last term in eqn 4 is a scalar term.

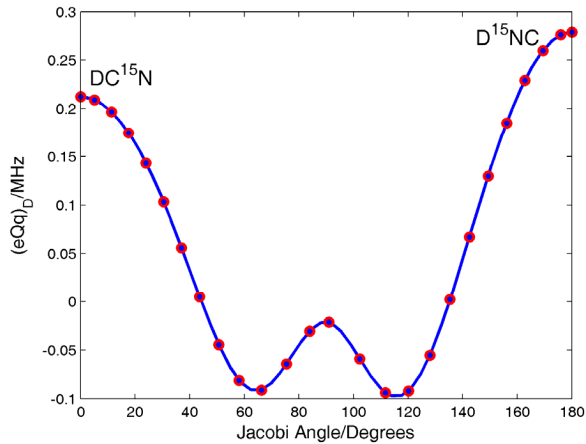

**Fig. 4** Dependence of the nuclear quadrupole coupling constant $(eQq)_D$ along the relaxed $DC^{15}N \rightleftarrows D^{15}NC$ isomerization path. The plot of $(eQq)_D$ as a function of the Jacobi angle is approximately symmetrical.

Like the acetylene ⇌ vinylidene isomerization system, which has been the subject of many theoretical studies[34-40], the HCN ⇌ HNC isomerization involves a periodic potential. Therefore, the complete periodic isomerization path can be constructed with only the information about the path from HCN to HNC by using permutation group operations in a local frame. Each resulting geometry was translated to a center of mass frame, and all Cartesian components as a function of the Jacobi angle were fit to a Fourier series. Finite differences were then used on the fitted geometries to align the molecule along an Eckart frame. Finally, the kinetic energy, potential energy, and quadrupole matrix elements were obtained from their Fourier series expansion coefficients, and the one-dimensional Hamiltonian in eqn (4) was diagonalized in a basis of complex exponentials

$$\left\{ \frac{e^{-mis}}{\sqrt{2\pi}}, \cdots, \frac{e^{-2is}}{\sqrt{2\pi}}, \frac{e^{-is}}{\sqrt{2\pi}}, \frac{1}{\sqrt{2\pi}}, \frac{e^{is}}{\sqrt{2\pi}}, \frac{e^{2is}}{\sqrt{2\pi}}, \cdots, \frac{e^{mis}}{\sqrt{2\pi}} \right\} \quad (7)$$

where $m$ is a positive integer.

## C. Progression of the Quadrupole Coupling Constant

The lowest 50 eigenenergies obtained from the internal coordinate Hamiltonian are presented in Fig. 5 (a) as horizontal lines superimposed on the HC$^{14}$N ⇌ H$^{14}$NC isomerization potential. Fig. 5 (b) illustrates the variation of the vibrationally-averaged $(eQq)_N$ coupling constants, $\langle (eQq)_N \rangle$, which are associated with each of the energy levels in Fig. 5 (a). Vibrational wavefunctions that are localized in the HC$^{14}$N and H$^{14}$NC potential wells are denoted by triangular and square symbols respectively. All assignments were based on the expectation values of cos($\theta$) until this metric became indeterminate at energies far above the transition state. For $n < 8$, $\langle (eQq)_N \rangle$ varies linearly since the vibrational wavefunction is localized in the HC$^{14}$N global minimum. After the HC$^{14}$N bending energy exceeds 5,378 cm$^{-1}$ (8 < $n$ < 52), $\langle (eQq)_N \rangle$ varies rapidly between two limits since the vibrational wavefunction alternates its localization between the local H$^{14}$NC minimum and the global HC$^{14}$N minimum. Once the bending energy surpasses 16,879 ($n >$ 52), the hydrogen migration becomes nearly a free rotation and $\langle (eQq)_N \rangle$ is approximately constant with a limiting value of approximately -4.3 MHz.

In contrast, the variation of vibrationally-averaged $(eQq)_D$ constants for the DC$^{15}$N ⇌ D$^{15}$NC isomerization is considerably different. Fig. 5 (c) shows the progression of $\langle (eQq)_D \rangle$ with assignments of DC$^{15}$N (triangles) and D$^{15}$NC (squares) vibrational energy levels. For low bending quanta, $\langle (eQq)_D \rangle$ is positive and follows a linear trend in both the DC$^{15}$N and D$^{15}$NC isomers. However, as the bending energy is increased, the $\langle (eQq)_D \rangle$ values for DC$^{15}$N and D$^{15}$NC intersect near $n = 34$ due to increased vibrational sampling of the $(eQq)_D = 0$ transition state. Above this energy, $\langle (eQq)_D \rangle$ oscillates between the DC$^{15}$N and D$^{15}$NC localized states before approaching a single constant value of 0.03 MHz at high excitation.

The numerical values of the bending frequency and $\langle eQq \rangle$

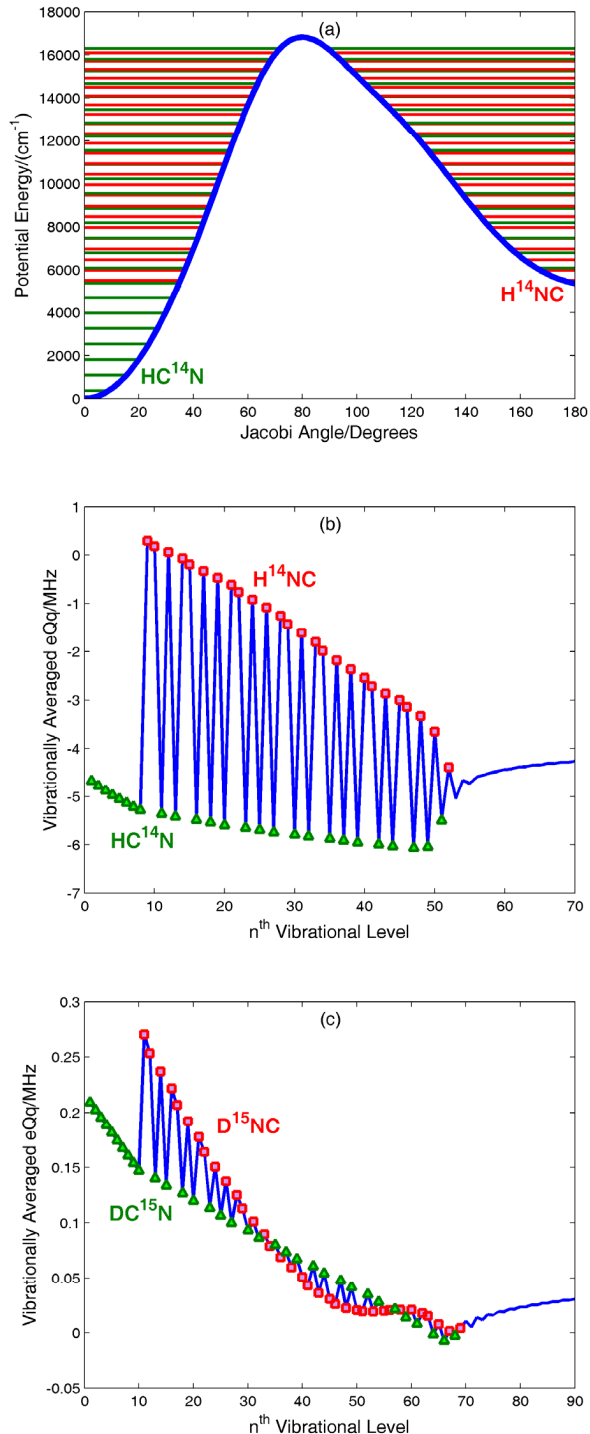

**Fig. 5** Progression of nuclear quadrupole coupling constants. Fig. 5 (a) depicts the 50 lowest vibrational eigenvalues for the HC$^{14}$N ⇌ H$^{14}$NC potential obtained at the CCSD(T)/cc-pCVQZ level of theory. Fig. 5 (b) shows vibrationally-averaged nuclear quadrupole coupling constants for vibrational states of HC$^{14}$N (triangles) and H$^{14}$NC (squares). Fig. 5 (c) depicts vibrationally-averaged coupling constants for DC$^{15}$N (triangles) and D$^{15}$NC (squares). Assignments for all plots are based on expectation values of cos($\theta$).

constants computed from the one-dimensional Hamiltonian are given in Tables 1 and 2 for HC$^{14}$N/H$^{14}$NC and DC$^{15}$N/D$^{15}$NC respectively. The data compiled in Table 1 is also compared with recent experimental values taken from millimeter-wave absorption measurements by Bechtel et al.[19] The agreement between the ab initio and experimental $\langle (eQq)_N \rangle$ values is excellent with only a 35 kHz deviation within the data set. Figs. 6 (a) and (b) present bending frequencies and $\langle (eQq)_N \rangle$ values as a function of vibrational excitation. Both the $HC^{14}N$ and $H^{14}NC$ frequencies obey a linear trend at low quanta with 710.2 cm$^{-1}$/($HC^{14}N$ bend quantum) and 494.4 cm$^{-1}$/($H^{14}NC$ bend quantum) respectively (the experimental data yields smaller slopes of 685.6 cm$^{-1}$/($HC^{14}N$ bend quantum) and 468.4 cm$^{-1}$/($H^{14}NC$ bend quantum)). Similarly, the vibrationally-averaged quadrupole coupling constants also obey a nearly linear trend with -0.078 MHz/($HC^{14}N$ bend quantum) and -0.140 MHz/($H^{14}NC$ bend quantum). Both the *ab initio* and experimental data for $H^{14}NC$ indicate that $\langle (eQq)_N \rangle$ changes sign near $n = 2$. The overall trends of the *ab initio* values are also in good agreement with the experimental findings of Bechtel et al. who measure coupling constants of -0.075 MHz/($HC^{14}N$ bend quantum) and -0.118 MHz/($H^{14}NC$ bend quantum). The close match between the *ab initio* and experimental values supports the assertion that the isomerization process is well-described by a single one-dimensional coordinate. Even more useful is the ability to identify specific vibrational levels based on predictions of the nuclear quadrupole hyperfine structure. The variation of $\langle (eQq)_N \rangle$ values as a function of vibrational excitation further demonstrates that nuclear quadrupole hyperfine structure can be used as a sensitive probe for characterizing localized $HC^{14}N$ and $H^{14}NC$ vibrational states.

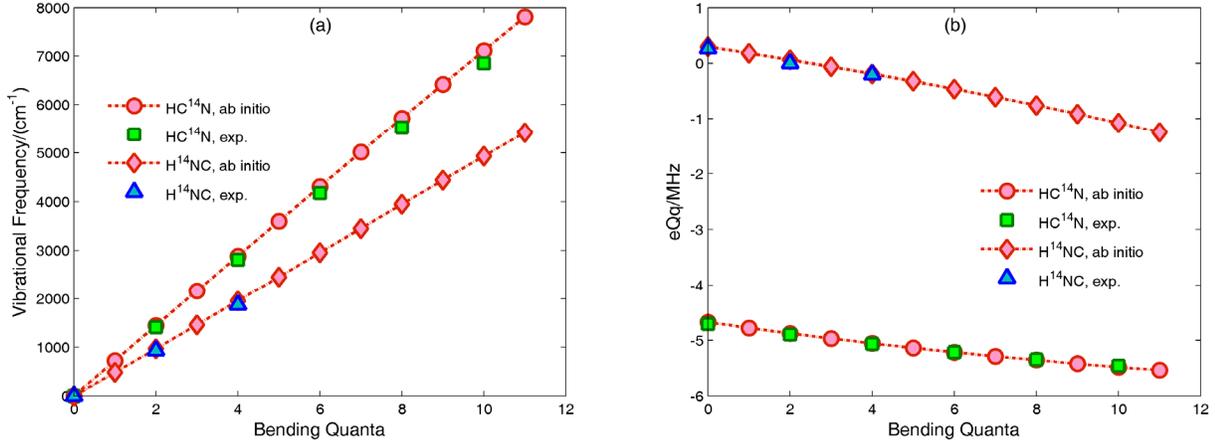

**Fig. 6** Comparison with experiment. Fig. 6 (a) depicts calculated and experimental ($l = 0$) vibrational frequencies for $HC^{14}N$ and $H^{14}NC$. Fig. 6 (b) shows calculated and experimental ($l = 0$) nuclear quadrupole coupling constants for $HC^{14}N$ and $H^{14}NC$.

**Table 1** Energies and nuclear quadrupole coupling constants of $HC^{14}N$ and $H^{14}NC$ vibrational eigenstates.

| Vibrational Level $(v_1, v_2, v_3)$ | | Ab initio Frequency (cm$^{-1}$) | Experimental Frequency[a] (cm$^{-1}$) | Ab initio $(eQq)_N$ (MHz) | Experimental $(eQq)_N$[a] (MHz) |
|---|---|---|---|---|---|
| $HC^{14}N$ | (0, 0, 0) | 0 | 0 | -4.6764 | -4.7084 |
| | (0, 1, 0) | 723 | — | -4.7796 | — |
| | (0, 2, 0) | 1445 | 1411 | -4.8771 | -4.8966 |
| | (0, 3, 0) | 2165 | — | -4.9691 | — |
| | (0, 4, 0) | 2882 | 2803 | -5.0561 | -5.0699 |
| | (0, 5, 0) | 3596 | — | -5.1382 | — |
| | (0, 6, 0) | 4308 | 4175 | -5.2157 | -5.2175 |
| | (0, 7, 0) | 5016 | — | -5.2888 | — |
| | (0, 8, 0) | 5720 | 5526 | -5.3578 | -5.3485 |
| | (0, 9, 0) | 6420 | — | -5.4229 | — |
| | (0, 10, 0) | 7116 | 6856 | -5.4843 | -5.4579 |
| | (0, 11, 0) | 7807 | — | -5.5423 | — |
| | (0, 12, 0) | 8493 | — | -5.5971 | — |
| | (0, 13, 0) | 9174 | — | -5.6490 | — |
| | (0, 14, 0) | 9848 | — | -5.6983 | — |
| | (0, 15, 0) | 10515 | — | -5.7452 | — |
| $H^{14}NC$ | (0, 0, 0) | 0 | 0 | 0.2961 | 0.2641 |
| | (0, 1, 0) | 479 | — | 0.1806 | — |
| | (0, 2, 0) | 966 | 927 | 0.0611 | 0.0451 |
| | (0, 3, 0) | 1458 | — | -0.0629 | — |
| | (0, 4, 0) | 1954 | 1874 | -0.1915 | -0.2066 |
| | (0, 5, 0) | 2452 | — | -0.3256 | — |
| | (0, 6, 0) | 2951 | — | -0.4651 | — |
| | (0, 7, 0) | 3450 | — | -0.6100 | — |
| | (0, 8, 0) | 3947 | — | -0.7608 | — |
| | (0, 9, 0) | 4441 | — | -0.9178 | — |
| | (0, 10, 0) | 4931 | — | -1.0812 | — |
| | (0, 11, 0) | 5417 | — | -1.2509 | — |
| | (0, 12, 0) | 5896 | — | -1.4266 | — |
| | (0, 13, 0) | 6367 | — | -1.6082 | — |
| | (0, 14, 0) | 6831 | — | -1.7948 | — |
| | (0, 15, 0) | 7284 | — | -1.9852 | — |

[a] Ref. 19

**Table 2** *Ab initio* energies and nuclear quadrupole coupling constants of $DC^{15}N$ and $D^{15}NC$ vibrational eigenstates. Vibrational assignments were based on expectation values of $\cos(\theta)$.

| Vibrational Level $(v_1, v_2, v_3)$ | | *Ab initio* Frequency (cm$^{-1}$) | *Ab initio* $(eQq)_D$ (MHz) |
|---|---|---|---|
| $DC^{15}N$ | (0, 0, 0) | 0 | 0.2089 |
| | (0, 1, 0) | 575 | 0.2021 |
| | (0, 2, 0) | 1149 | 0.1952 |
| | (0, 3, 0) | 1721 | 0.1883 |
| | (0, 4, 0) | 2291 | 0.1814 |
| | (0, 5, 0) | 2859 | 0.1745 |
| | (0, 6, 0) | 3425 | 0.1676 |
| | (0, 7, 0) | 3989 | 0.1607 |
| | (0, 8, 0) | 4550 | 0.1539 |
| | (0, 9, 0) | 5108 | 0.1470 |
| | (0, 10, 0) | 5663 | 0.1402 |
| | (0, 11, 0) | 6215 | 0.1334 |
| | (0, 12, 0) | 6764 | 0.1266 |
| | (0, 13, 0) | 7310 | 0.1198 |
| | (0, 14, 0) | 7852 | 0.1131 |
| | (0, 15, 0) | 8391 | 0.1064 |
| $D^{15}NC$ | (0, 0, 0) | 0 | 0.2704 |
| | (0, 1, 0) | 370 | 0.2534 |
| | (0, 2, 0) | 744 | 0.2372 |
| | (0, 3, 0) | 1122 | 0.2216 |
| | (0, 4, 0) | 1502 | 0.2066 |
| | (0, 5, 0) | 1884 | 0.1920 |
| | (0, 6, 0) | 2266 | 0.1777 |
| | (0, 7, 0) | 2649 | 0.1640 |
| | (0, 8, 0) | 3032 | 0.1506 |
| | (0, 9, 0) | 3415 | 0.1376 |
| | (0, 10, 0) | 3796 | 0.1250 |
| | (0, 11, 0) | 4175 | 0.1128 |
| | (0, 12, 0) | 4552 | 0.1010 |
| | (0, 13, 0) | 4927 | 0.0897 |
| | (0, 14, 0) | 5298 | 0.0790 |
| | (0, 15, 0) | 5665 | 0.0689 |

## V. Conclusions

This work has examined the nuclear quadrupole hyperfine structure in the $HC^{14}N/H^{14}NC$ and $DC^{15}N/D^{15}NC$ systems and demonstrated their diagnostic utility in characterizing vibrational localization and isomerization. In order to obtain quantitative agreement with experiment, CCSD(T) calculations with large core valence basis sets were needed to describe the electric field gradients at the nuclei of interest. Using these methods for the $HC^{14}N \rightleftarrows H^{14}NC$ isomerization, it was found that the $^{14}N$ quadrupole coupling constant is large in the vicinity of the $HC^{14}N$ potential well, making it an ideal system for millimeter wave spectroscopic studies. On the other hand, for the $DC^{15}N$ and $D^{15}NC$ isotopomers, small coupling constant constants were calculated, which arise from the nearly symmetrical chemical environment experienced by the deuteron nucleus during the isomerization process. In addition, the NBO population analysis for the $HC^{14}N/H^{14}NC$ system supports the simple model of Townes and Dailey which maintains that hyperfine structure arises from the unequal electron distribution among the $p_x$, $p_y$, and $p_z$ orbitals of the quadrupolar $^{14}N$ nucleus.

Furthermore, since hyperfine structure is sensitive to minor changes in the geometry, it is not sufficient to study only the HCN/HNC minima, and vibrational averaging of the entire isomerization path is necessary. The direct comparison between the vibrationally averaged $(eQq)_N$ values with experimental data demonstrates that the progression of hyperfine structure can be accurately predicted by a one-dimensional, internal coordinate Hamiltonian. Currently, the hyperfine structure of vibrationally-excited states of $DC^{15}N/D^{15}NC$ have not been studied, and experimental measurements of these $(eQq)_D$ values would be extremely valuable as a check on the accuracy of the theoretical calculations. Finally, nuclear quadrupole hyperfine interactions provide another useful electronic property to detect the onset of isomerization. Unlike the dipole moment, $\mu$, where the experimental observable is the *magnitude*, $|\mu|$, and not its sign ($|\mu| \sim 3$ Debye for both HCN and HNC), a hyperfine splitting measurement can determine in which potential well the vibrational wavefunction is localized.

## Acknowledgements

The author is grateful for helpful discussions with Prof. Robert W. Field, Dr. Hans A. Bechtel, and Adam H. Steeves. This work was partially supported by the National Center for Supercomputing Applications under grant number TG-CHE070084N and utilized the NCSA Cobalt SGI Altix System. Sandia is a multiprogram laboratory operated by Sandia Corporation, a Lockheed Martin Company, for the United States Department of Energy's National Nuclear Security Administration under contract DE-AC04-94AL85000.